\documentclass[12pt]{iopart}

\usepackage{graphicx}
\usepackage{bm}
\begin{document}

\title[Influence of the substrate lattice structure on the formation of QWS]{Influence of the substrate lattice structure on the formation of Quantum Well States in thin In and Pb films on silicon.}

\author{J H Dil$^{1,*}$, B H\"{u}lsen$^{1}$, T U Kampen$^{1}$, P Kratzer$^{2}$, and K Horn$^{1}$}

\address{$^1$Fritz-Haber-Institut der Max-Planck-Gesellschaft, Faradayweg 4-6, 14195 Berlin, Germany\\ $^2$Fachbereich Physik, Universit\"{a}t Duisburg-Essen, 47048 Duisburg, Germany\\ $^{*}$Present address: Physik-Institut, Universit\"{a}t Z\"{u}rich, Winterthurerstrasse 190, 8057 Z\"{u}rich, Switzerland}
\ead{jan-hugo.dil@psi.ch}

\begin{abstract}
The substrate lattice structure may have a considerable influence on the formation of quantum well states in a metal overlayer material. Here we study  three model systems using angle resolved photoemission and low energy electron diffraction: indium films on Si(111) and indium and lead on Si(100). Data are compared with theoretical predictions based on density functional theory. We find that the interaction between the substrate and the overlayer strongly influences the formation of quantum well states; indium layers only exhibit well defined quantum well states when the layer relaxes from an initial face-centered cubic to the bulk body-centered tetragonal lattice structure. For Pb layers on Si(100) a change in growth orientation inhibits the formations of quantum well states in films thicker than 2 ML.

\end{abstract}


\section{Introduction}
Ultrathin metal films play an increasingly important role in many technologies and are the subject of intense scientific research. The formation of quantum well states (QWS) in such films when their thickness becomes comparable to the electron coherence length is nowadays well understood\cite{Chiang}. The electronic and structural properties of such systems have yielded many unexpected results, such as the formation of magic heights\cite{Budde}, oscillations in the surface reactivity\cite{Aballe}, sign and magnitude of the Hall coefficient\cite{Jaloch}, and the formation of extra sub-bands\cite{Speer}. In many instances, research on quantum well states has contributed to a better understanding of the physics of low-dimensional structures. Furthermore the evolution of quantum well states can be used as a probe for the structure of thin metal films\cite{Luh1}.\newline
With smaller film thickness, the volume-to-surface ratio decreases and the influence of the interface between the metal overlayer and the substrate increases. In the present paper, the influence of the lattice structure of the substrate on the formation of quantum well states in the metal overlayer will be discussed. Indium films on Si(111) and Si(100) are chosen as an example of heteroepitaxy where even the crystal symmetry of the metal and the substrate is different. The formation of QWS in Pb on Si(111) has been discussed previously\cite{Upton}, and is used in this work as a reference. We extend these studies to Pb on Si(100) to examine the influence of an energetically unfavourable growth direction on the metal overlayer imposed by the substrate. For the systems dealt with here we find that the substrate lattice structure may have a profound effect on the formation of QWS, and that in some cases confinement is hindered by such effects.

\section{Experimental and computational set-up}
The experiments were carried out in a stainless steel ultrahigh vacuum chamber with a base pressure of $1 \times 10^{-10}$ mbar. The measurements were performed at the 10 m normal incidence monochromator on the U125/2 undulator at BESSY II. Data were acquired using a Phoibos 100 electron energy analyzer (Specs Gmbh) equipped with a CCD detector, thus allowing for the simultaneous detection of emission angle and electron energy. The line scans presented in this work were obtained by cutting through the images without further angular integration. Under the typical experimental conditions used here the energy and angular resolution are better than 80 meV and $0.2^{\circ}$, respectively.\newline 
The Si samples were cleaned by repeated flash annealing, after which the cleanliness was checked by the observation of a sharp LEED pattern, and the presence of the contamination sensitive surface states. Pb was deposited from a water-cooled Knudsen cell on the sample held at 100 K, and the measurements were performed at the same temperature. At this temperature, no well ordered In layers form on Si(111), therefore the experiments involving In were performed at a sample temperature of 50 K using liquid helium for cooling of our flow cryostat. The deposition rate was initially calibrated using a quartz microbalance and further refined based on the development of the quantum well states as a function of coverage.\newline
Band structure calculations were performed within the generalized-gradient approximation\cite{Perdew} to density functional theory (DFT), using the highly accurate all-electron full-potential linearized augmented plane wave (FP-LAPW) method implemented in the Wien2k code\cite{Schwarz}.
\begin{figure}[htb]
\begin{center}
\includegraphics[width=0.9\textwidth]{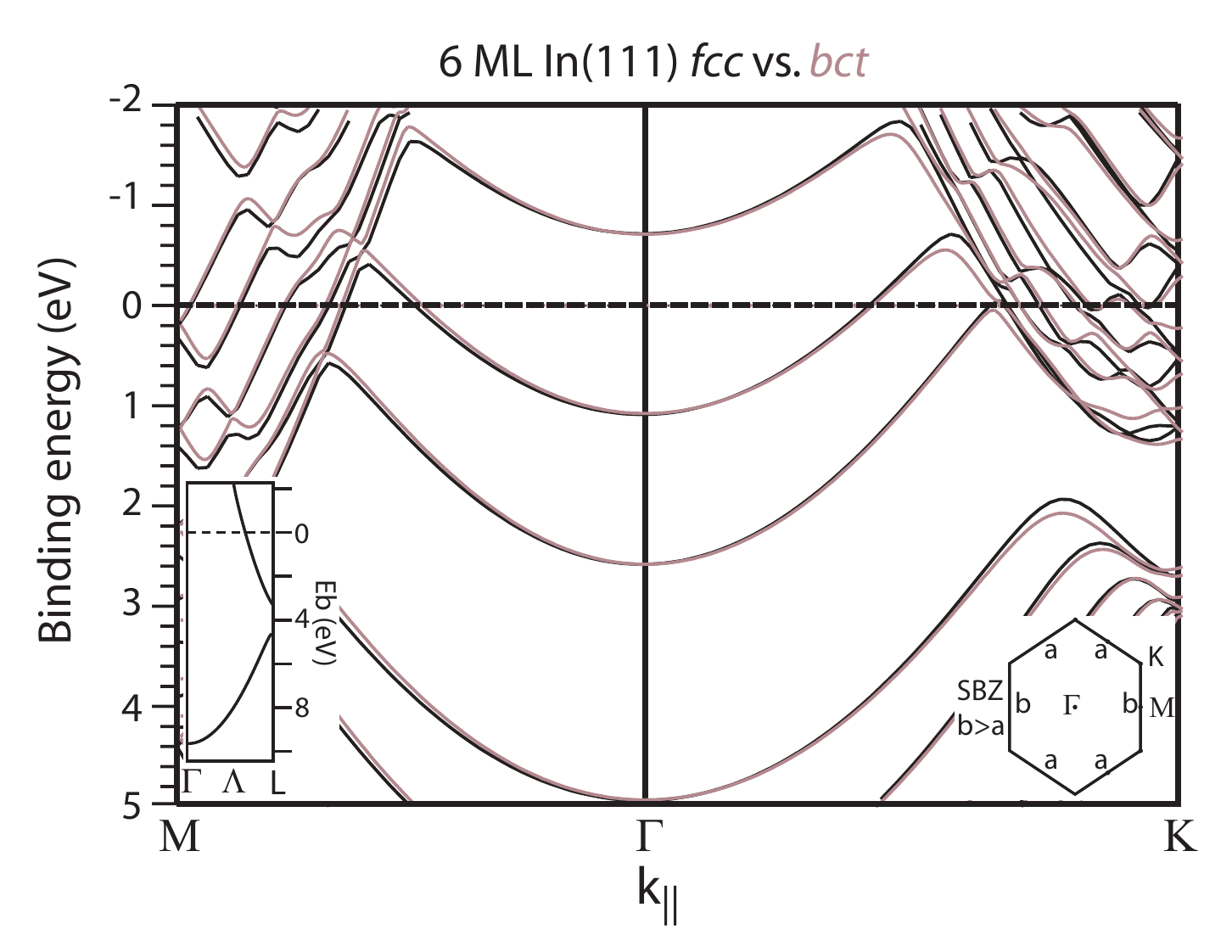} 
\caption{{Comparison between the calculated band structure of a 6 ML thick freestanding (111) film of \textit{fcc} (black line) and \textit{bct} (grey line) indium. The inset shows the predicted surface Brillouin zone for \textit{bct} indium where $a$ and $b$ are the reciprocal lattice spacing.}}
\label{Fig1}
\end{center}
\end{figure}

\section{Results and discussion}
Bulk indium crystallizes in the centred tetragonal lattice structure, also called body-centered tetragonal (\textit{bct}). The indium lattice is contracted about 3 \% in one direction compared to the face centred cubic (\textit{fcc}) structure. For \textit{bct} films grown in the [111] direction the surface Brillouin zone (SBZ) is shown in the inset of Figure \ref{Fig1}. How the difference in lattice structures between Si and In affects the growth of thin In layers will be discussed in the second section of this paper concerned with interface structure effects. The relatively small lattice distortion away from the higher symmetry \textit{fcc} structure has only a limited effect on the quantum well states, as can be observed in Figure \ref{Fig1}, where DFT band structure calculations for \textit{fcc} and \textit{bct} free-standing 6 monolayer (ML) thick indium slabs are displayed. The conduction band of indium has mainly $5p$ character. The set of upward-dispersing parabolae around the zone centre represent the states originating from the confinement of the $5p_z$ derived band in the quantum well, while the large number of states dispersing down towards the edge of the SBZ are due to quantization of the $5p_{x,y}$ derived bands. As is directly visible from the figure the difference in the QWS energies between the two structures at the zone centre is negligible; small (tens of meV) differences only occur towards the edge of the Brillouin zone. These differences are too small to provide a distinction between either structure through ARPES measurements in view of the width of the bands as shown below.

\subsection{ARPES of In films on Si(100)}
\begin{figure}[htb]
\begin{center}
\includegraphics[width=1\textwidth]{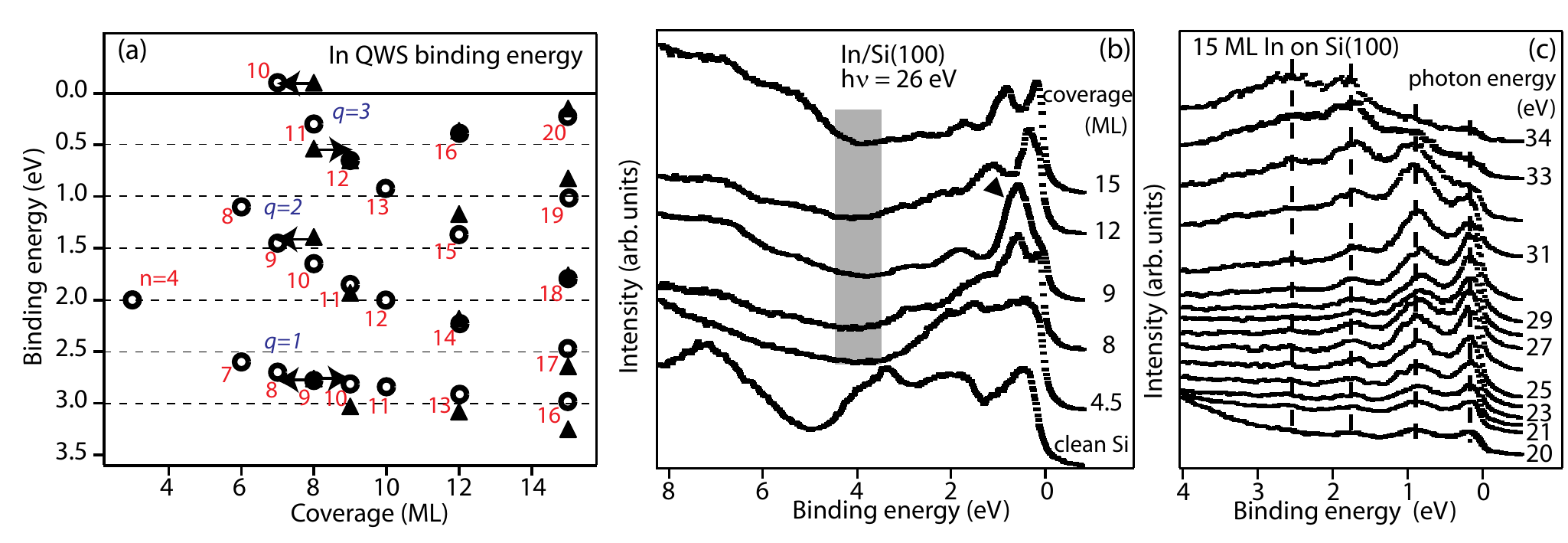} 
\caption{{(a) Comparison between the measured QWS binding energies for In/Si(100) (triangles) and the calculated QWS binding energies of freestanding In layers (open circles) as a function of thickness. The quantum number $n$ and the reduced quantum number $q$ are indicated. The arrows represent the observation that for a nominal coverage of 8 ML, the layer consists of a combination of 7 and 9 ML thick regions. (b) EDCs of different coverages of In on Si(100) obtained at normal emission at a photon energy of 26 eV, the shaded area represents the fundamental bandgap in bulk In. (c) EDCs for 15 ML of In on Si(100) obtained at normal emission as a function of photon energy. The dashed lines are added as a guide to the eye to indicate the absence of dispersion in the k$_{\bot}$ direction.}}
\label{Fig2}
\end{center}
\end{figure}

The calculated energies for the quantum well states at the parallel wave vector $k_{||} = 0$ in the topmost, partly occupied bulk In valence band\cite{Ashcroft} are displayed as empty circles in Figure \ref{Fig2}(a): at a thickness of 3 ML only one QWS at a binding energy of 2.0 eV is occupied, at 6 ML there are two, and at 9 ML a third QWS is occupied. This is related to the shape of this band and its Fermi level crossing at about $\frac{1}{3}$ along the $\Gamma -$L line in the 3D band structure as shown in the inset of Figure \ref{Fig1}\cite{Neuhold}.\newline
Quantum well states for a single thickness as indicated in the plot are numbered starting from the deepest state located in the fully occupied valence band which has a binding energy of approximately 10 eV. The QWS in indium layers of different thickness that have a binding energy close to each other only differ in their quantum number by one; the quantum number $n$ of the state at 1.1 eV in the 6 ML film is 8, for the state at 1.45 eV in the 7 ML film $n = 9$, and for the state at 0.3 eV in the 8 ML film $n = 11$. Based on this a reduced quantum number $q$ can be defined for the In QWS as $q = n - N$, where $N$ is the layer thickness, resulting in branches of constant $q$ that move towards higher binding energies; they are also given in Figure \ref{Fig2}(a).\newline 
The calculated quantum well state energies are compared with the experimental data derived by evaluating the binding energies in energy distribution curves as shown in Figure \ref{Fig2}(b) for normal emission photoemission from clean Si(100) and several depositions of In on Si(100), obtained at a photon energy of 26 eV. For coverages of 8 ML and more, several features are present that are not observed for the clean Si(100) surface. These lines show a strong dependence on the amount of indium deposited, which directly suggests that the features are derived from QWS. For the 4.5 ML thick film it is hard to distinguish any clear QWS and to determine their energies, although such features appear to be emerging. This is most likely caused by the fact that, at this coverage, there is no homogeneous layer, but rather islands of various heights, resulting in many states very close to each other. The shape of the photoemission background upon which the QWS features are placed at this coverage does, however, strongly resemble the one observed for higher coverages. Here sharp features, identified as QWS, are observed down to around 3.5 eV below the Fermi level. Between 3.5 and 4.5 eV there is a gap in the spectra, which corresponds to the bandgap in the $\Lambda$ -direction for bulk In\cite{Ashcroft}. Below this gap, some broad lines can be observed. These are expected to originate from QWS in the lower valence band (see bulk band structure in the inset of Figure \ref{Fig1}). In contrast to the upper valence band, for the lower valence band a new QWS is occupied for each additional layer because the band stretches across the entire Brillouin zone, therefore the states are close together. From this data it is clear that QWS are also formed in the lower valence band of indium; however, since these states, being much further away from the Fermi level, show strong lifetime broadening and are not well resolved due to an overlapping of states, they will not be discussed any further here.\newline
Some of the QWS peaks in Figure \ref{Fig2}(b) show a slight shoulder on the higher binding energy side, as for example  the peak closest to the Fermi level in the 9 ML thick film. This belongs to the 10 ML thick film, as indicated by the triangle. This indicates that the coverage is slightly more than 9 ML, and therefore areas with 10 ML coverage exist. Such features will be taken into account in the evaluation of the data in order to more accurately determine the energies of the main features in the spectra. The binding energies for the most prominent QWS in thin In layers on Si(100) are indicated in Figure \ref{Fig2}(a) as a function of coverage by solid triangles, while the QWS energies obtained from DFT calculations for free-standing indium films are indicated as open circles.\newline
For an 8 ML indium film, it is clear that the experimental QWS energies do not match the calculated results. From a comparison with the binding energies obtained by DFT, the QWS measured for a 8 ML thick film can be ascribed to a combination of a 7 and 9 ML thick film. This means that the 8 ML film is not formed at all$!$. A possible explanation for this is the influence of quantum size effects (QSE) on film growth, similar to those responsible for the formation of stable and unstable islands heights in other systems\cite{Smith, Luh2, Otero, Dil1}.The QWS expected for an 8 ML thick film is very close to $E_F$ which significantly increases the total energy of the system. When the QWS observed after deposition of 8 ML of In are properly assigned, as discussed above and indicated by the arrows, the measured energies match the calculated ones very well. We therefore conclude that indium grows in a layer-by-layer fashion for coverages of 9 ML or more. It seems that below this coverage, a mixture of layer heights is formed, which corresponds to the data for both the 4.5 and 8 ML depositions. This interpretation is analogous to the "inverse Stranski-Krastanov growth" observed for Ag on GaAs(110)\cite{Smith} where, after initial island growth, from a certain coverage onward atomically smooth layers are formed.\newline
Due to the confinement in the direction normal to the surface, QWS show no dispersion in this direction with the wavevector component $k_\bot$, and their binding energies should remain constant when changing the photon energy. Figure \ref{Fig2}(c) shows the photon energy dependence of the QWS in a 15 ML thick indium film on Si(100), for photon energies between 20 and 34 eV. The individual spectra are obtained by taking a slice through an "energy versus angle" image at normal emission, and normalizing the intensities to the photon flux. Two clear observations can be made from this figure. First, the QWS show no noticeable dispersion as a function of photon energy, confirming the confinement of wave functions in the direction normal to the film surface. Secondly, there are strong variations in the intensity of the QWS as a function of photon energy. From a photon energy of 20 eV onwards, the intensity of the QWS continues to increase up to a photon energy of around 29 eV. Above this photon energy, a behaviour is observable that shows strong similarities to results for QWS in well ordered Pb films\cite{Dil2}: the weight of the QWS intensities shifts away from the Fermi level, and follows the dispersion of the bulk band. This can be understood by considering the fact that the wave function in the quantum well consists of the rapidly oscillating Bloch wave function of the (bulk) metal convoluted with the quantum well state envelope wave function. Hence the QWS show some bulk-like character in their photoemission cross section, which is dominated here by the matrix element for transitions form the initial to the final state band for bulk indium. From this plot an optimum photon energy that combines a high cross section with a broad energy window in the analyser can be determined. This optimum is located around 26 eV, hence this is the photon energy that was used in most of the experiments involving thin indium layers.
\begin{figure}[htb]
\begin{center}
\includegraphics[width=0.9\textwidth]{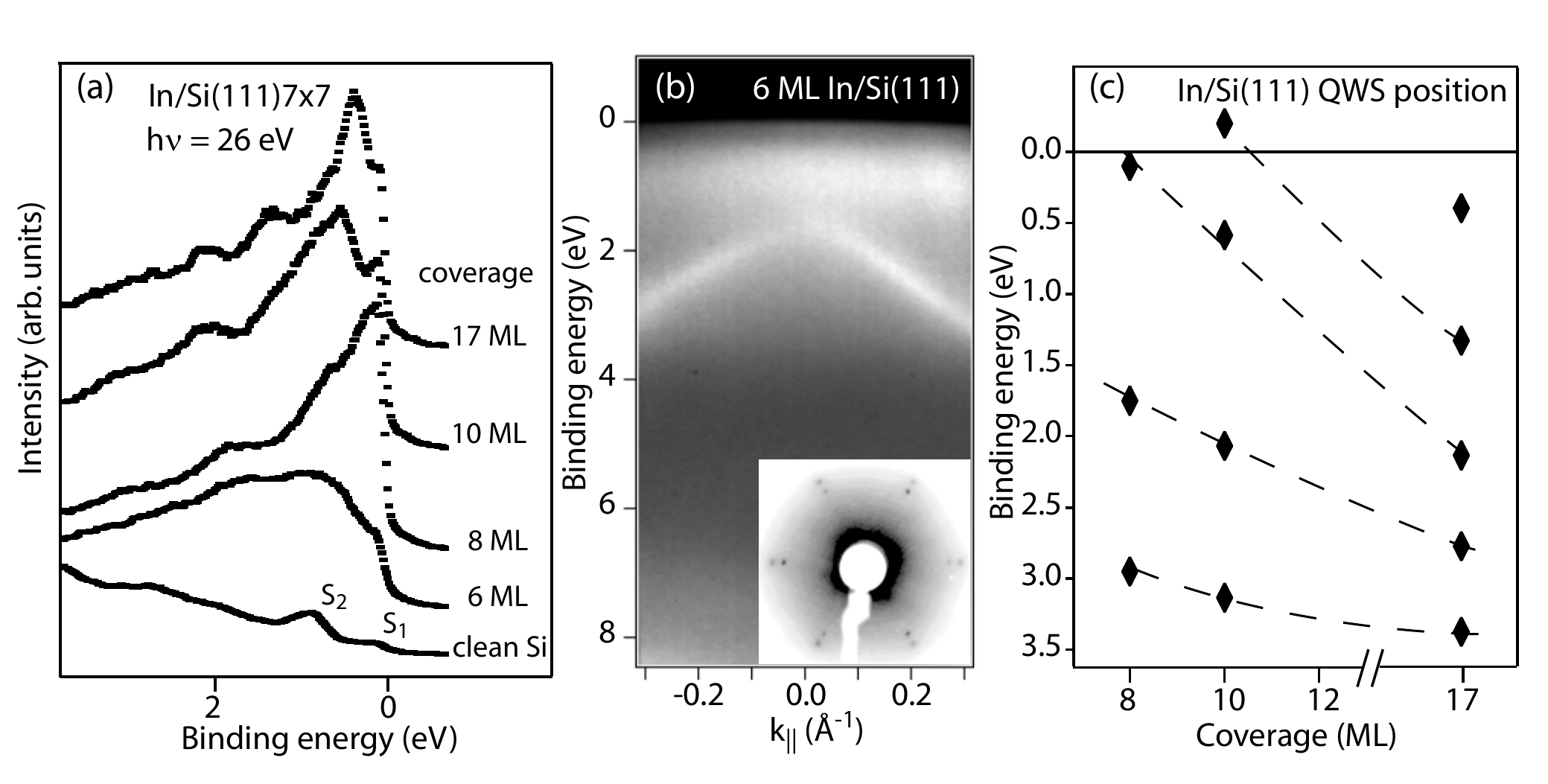} 
\caption{{(a) EDCs for In deposited on Si(111)$7\times7$ at 50K, obtained at normal emission with a photon energy of 26 eV. (b) Energy vs. momentum image for 6 ML In on Si(111)$7\times7$, (inset) corresponding LEED image. (c) QWS binding energy position as a function of coverage for In on Si(111)$7\times7$. The dashed lines correspond to QWS with the same quantum number $q$.}}
\label{Fig3}
\end{center}
\end{figure}
\subsection{In films on Si(111) $7 \times 7$}
For indium deposited on the reconstructed Si(111) $7 \times 7$ surface, the measured energy distribution curves are shown in Figure \ref{Fig3}(a). The bottom spectrum is for clean Si(111), where two surface states can be observed; the Si-adatom-derived $S_1$ state at a binding energy of 0.2 eV, and the restatom-derived $S_2$ state at EB = 0.8 eV\cite{Uhrberg}, as indicated in the plot. For a six layer thick film of indium, the Fermi edge is well developed, but only broad features are present in the valence band. The image from which this spectrum has been extracted (Figure \ref{Fig3}(b)), shows the downward dispersing silicon valence band. Directly above this sharp feature, a broad band can be observed that shows almost no in-plane dispersion. This corresponds to the broad peak that occurs in the EDC for a 6 ML film in Figure \ref{Fig3}(a). The fact that the silicon valence band is still clearly visible shows that, just as on Si(100), indium grows in islands for these low coverages. This is confirmed by the LEED image for this coverage(inset in Figure \ref{Fig3}(b)). The inner spots, originating from the Si(111) surface, are more intense than the indium spots, which due to the difference in lattice spacing occur just outside the Si spots. Island-like growth for low coverages of In on Si(111) has also been observed in STM studies by Altfeder \etal \cite{Altfeder}, which showed that the islands are slightly elongated in the direction along the step edges due to reduced diffusion perpendicular to the steps. The STM data further show that these islands are larger than 100 nm in all directions parallel to the surface.\newline
For coverages of 8 ML or more, it appears that the islands close to form a smooth layer, because individual QWS can be clearly resolved. Moreover these states show a strong dependence on the total coverage. The gap just below 4 eV binding energy, indicated by the dark region in Figure \ref{Fig3}(b) can be clearly seen, as in In/Si(100). A major difference, however, is the intensity difference between the QWS within approximately 1 eV below the Fermi level and the deeper lying states. For In/Si(100), the QWS intensity showed a gradual decrease away from EF, whereas here the transition is very abrupt. This cannot be a matrix element effect, because both data sets have been recorded at the same photon energy; and because both films grow in the [111] direction. This discrepancy is therefore attributed to interface effects, as explained below. \newline
Figure \ref{Fig3}(c) shows the binding energy position of the primary QWS extracted from the EDC in Figure \ref{Fig3}(a). It can be directly concluded that the appearance of QWS in  indium layers grown on Si(111) follows the same rules as predicted by DFT and observed for In/Si(100). For every three additional monolayers deposited, one additional QWS is formed in the upper valence band because of the crossing of the Fermi level about $\frac{1}{3}$ along the distance from the Brillouin zone boundary.
\begin{figure}[htb]
\begin{center}
\includegraphics[width=0.9\textwidth]{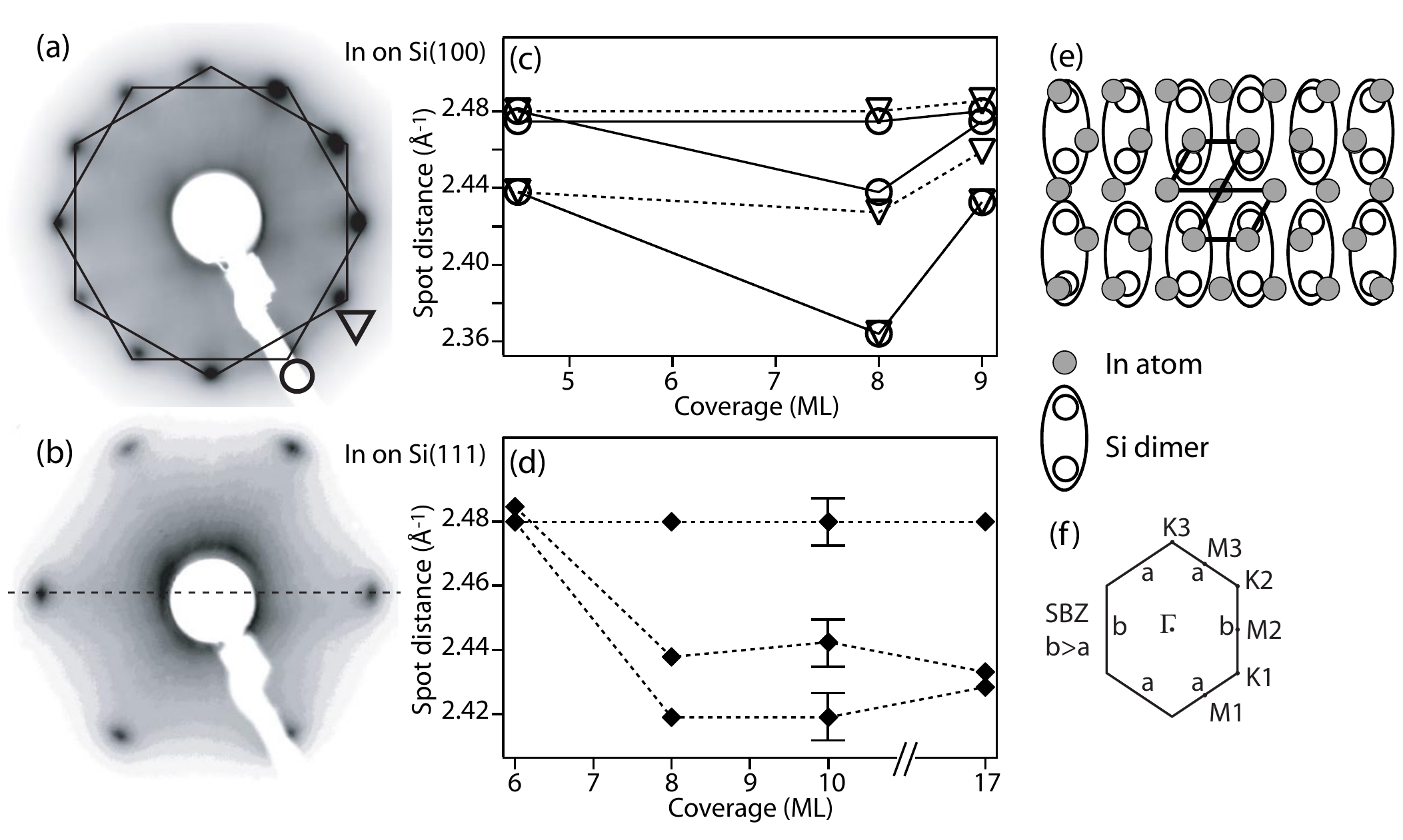} 
\caption{{(a and b) LEED patterns for a 10 ML thick layer of In grown on Si(100) (a) and on Si(111) (b). The dashed line in (b) indicates along which direction intensity traces are obtained. The hexagons and symbols in (a) refer to the symbols used in (c). (c and d) LEED spot distances as a function of In coverage on (c) Si(100) and (d) Si(111) demonstrating the presence of \textit{bct} and \textit{fcc} In layers. The different symbols in (c) represent the different domains and the lines connect the geometrical orientations that belong together. The error bars for 10 ML in (d) are representative for all the data points in (c) and (d). (e) Proposed lattice structure for one domain of In on Si(100). (f) (inset) \textit{bct} surface Brillouin zone with $b > a$.}}
\label{Fig4}
\end{center}
\end{figure}
\subsection{Structural properties of In films on Si(100) and Si(111)}
The detailed structure of the metal-semiconductor interface, and of the first few layers of the metal film can be of substantial influence on the occurrence of quantum well states from the metal bands. We showed that from a coverage of approximately 9 ML, indium grows in a layer-by-layer fashion, both on Si(100) and on Si(111). However, a major difference exists between the growth mode on the two different substrates as seen in Figure \ref{Fig4} where LEED patterns from a 10 ML thick film of indium on Si(100) (a) and a film of identical thickness on Si(111) (b) are shown. The most striking distinction between the two patterns lies in the number of spots: 6 for the layer on Si(111) and 12 for the layer on Si(100). From the comparison of QWS energies to DFT results, and from the hexagonal LEED pattern, we conclude that on both substrates the layers grow in the [111] direction. Since the Si(100) surface consists of two domains rotated 90$^\circ$ with respect to each other, the indium overlayer also grows in two domains. This explains why there are 12 diffraction spots for In/Si(100) and 6 for In/Si(111). From the sharpness of the spots the good crystalline quality of the layer, which was expected from the well defined QWS shown in Figures \ref{Fig2} and \ref{Fig3}(a), can be confirmed, at least inasmuch as reflected in the occurrence of quantum well states.\newline
Bulk indium has a body-centred tetragonal crystal structure, which means that when cutting to obtain a [111] plane, the surface Brillouin zone has the shape of a slightly elongated hexagon as shown in Figure \ref{Fig4}(f) , where for clarity the distortion from the \textit{fcc} hexagon is exaggerated. In real space the distance $a$ is 3.8\% larger than the interatomic distance $b$; in the SBZ this is inverted. This distortion of the perfect hexagon should also occur in LEED; the pattern should appear elongated in one direction. For the LEED images presented in Figure \ref{Fig4}, this elongation is not directly obvious. In order to obtain a quantitative result, a numerical analysis of the LEED pattern was carried out using the following routine: along a horizontal axis, cutting exactly through two spots as indicated by the dashed line in Figure \ref{Fig4}(b), an intensity trace is extracted from the image. The peaks in this spectrum are then fitted with a Gaussian function in order to deduce the exact peak positions. The use of spacings between two spots in the subsequent analysis, instead of their absolute position, reduces the influence of errors produced by a slight angular misalignment of the sample in front of the LEED instrument. The image is then rotated around the centre of the LEED pattern to obtain the next two spots along the horizontal axis, taking extra precaution not to distort the image in the process, and the analysis is repeated. Following this procedure, for In/Si(111) three distances, and for In/Si(100) six distances are obtained. Repeating this routine for several coverages results in a spot distance versus coverage plot. Figure \ref{Fig4}(c) shows such a plot for In/Si(100), wherein the different symbols show the two domains as indicated by the hexagons and symbols in Figure \ref{Fig4}(a). The results for In/Si(111) are shown in Figure \ref{Fig4}(d); in both (c) and (d), lines connect those geometrical orientations (i.e. $\bar{\Gamma}-\bar{\mathrm{K_1}},\bar{\mathrm{K_2}},\bar{\mathrm{K_3}}$) that belong together.\newline
From the fact that for the indium films on Si(111), the spot distances for a coverage of 6 ML are the same in all three directions, we conclude that the In layer initially grows in a face-centred cubic structure. Since this structure for indium is only known in a high-pressure phase\cite{Simak}, this is an example of pseudomorphic growth. For higher coverages, the layer relaxes towards a body-centred tetragonal crystal structure, indicated by the observation that two distances in reciprocal space are shorter than the other. For the 17 ML thick layer, the difference between the long and short axis is approximately 2\%. Since for bulk In this value is 3\%, the layers are not expected to be fully relaxed towards the bulk lattice at this film thickness. The observation of a phase transition from \textit{fcc} to \textit{bct} is in line with STM and diffraction results for In/Si(111)-Pb-$(\sqrt3\times\sqrt3)$\cite{Chen}, where it has to be noted that our experiments have been performed at lower temperatures to enable the growth on the $(7\times 7)$ reconstructed surface.\newline 
Although the calculated data in Figure \ref{Fig1} suggest that the lattice structure should be of little influence on the electronic structure, in the In layers with an \textit{fcc} structure, no clear QWS are formed, and the energy versus momentum image in Figure \ref{Fig3}(b) only shows a broad non-dispersing feature just above the silicon valence band maximum. We speculate that there is a connection between non-dispersive features in the quantum well spectra and the possibility to map the interface below the metal film by STM. For example, in a recent STM study\cite{Altfeder} it was shown that the underlying Si(111)7x7 surface could be imaged through indium islands with a height of as much as 7 ML. Because STM is a method in which electrons with k$_{||}$ over a wide range may contribute, structural information buried below a metal layer cannot normally be obtained. This limitation is even valid for thin metal layers, where an electron can maintain its phase information over the distance travelled through the film. Because the electronic states observed for \textit{fcc} indium films are non-dispersing, they may provide a pathway through which an electron transported to the surface may retain local structural information from the interface. To understand this, we first note  that the tip, when imaging, extracts or injects electrons with a sharply defined energy. If the same state has a constant energy over a broad k$_{||}$ range, it is possible to describe the emitted electrons by a narrow wave packet in real space (broad in k-space) which acts as a local probe for the transmission through the interface. This interpretation of the STM data is supported by the fact that for indium films thicker than 7 ML (for which dispersing states develop) the In/Si interface is no longer visible in STM. In the photoemission study presented here, this is confirmed by the formation of QWS with a free-electron-like dispersion. This means that electrons with a sharply defined energy must also have a narrow range of k-vectors, and thus form a broad wave front in real space that averages over the local interface structure. The same physical principle may be responsible for the observation of the Si(111)7x7 surface reconstruction through Pb layers of 100 \AA\ thickness\cite{Altfeder2}, although in that system the lack of dispersion in the QWS sub-bands was suggested to originate from electron localization parallel to the surface which is still present for Pb layers of more than 20 ML thickness\cite{Dil3}.\newline
The limiting factor for the formation of QWS is not the \textit{fcc} lattice structure of the In films itself, since the DFT calculations indicate very little influence of the crystal structure (Fig. 1), nor is it likely to be the island-like growth because these islands are larger than 100 nm\cite{Altfeder}. In a recent theoretical approach on layer growth where the substrate was included in the analysis it was suggested that thin In layers are not allowed to form due to stress induced by the substrate\cite{Prieto}. The reason why QWS do not form could probably be elucidated by STM and X-ray diffraction studies. Based on the results presented here we can state that the transition to well ordered crystalline films that can accommodate QWS coincides with the transition from a \textit{fcc} to \textit{bct} lattice structure, but that the mechanism responsible for this effect is not yet clear.\newline
From the general trend in the spacing of the LEED spots as a function of coverage it is clear that the growth mode of indium on Si(100) differs from that on Si(111). First, on Si(100) not all spot distances are identical for thinner layers, suggesting that the initial growth is \textit{bct} and not \textit{fcc}-like. The asymmetry in spot distances remains similar for higher coverages, indicating that the thinner layers are relaxed in a manner comparable to the thicker ones. For In/Si(111) it was seen in the LEED pattern that in the \textit{bct} growth region, one distance is larger than the two others; this would also be expected from the bulk lattice structure. Measurements for In/Si(100) yield a different result; as explained below the asymmetry for the two different domains is opposite to each other. The triangular markers for domain 1 show one distance being larger than the others. On the other hand, domain 2, represented by the circles, shows one spot spacing being shorter than the other two. Superposition of the signal from these two domains rotated by 90$^\circ$ with respect to each other results in the oval shape of the whole LEED pattern in Figure \ref{Fig4}(a). The variations in distance and the general oval shape of the LEED pattern are also confirmed by the fact that, in order to rotate the image to horizontally align the next spots, variations in the rotation angle of as much as 3$^\circ$ are needed. For a 180$^\circ$ rotation, these variations cancel each other, i.e. rotating by exactly 180$^\circ$ maps the pattern onto itself. From a comparison to LEED images obtained for the clean Si(100) substrate directly before indium deposition, it can be concluded that the "odd" distance (i.e. the distance shorter/longer than the other two) is aligned parallel to the dimer rows of the substrate. This alignment, perpendicular to the dimers, is then the only way in which the square symmetry of the substrate can be partly matched by the \textit{bct} symmetry of the overlayer as schematically indicated in Figure \ref{Fig4}(e). The asymmetry between the growth on the different substrate domains is most likely induced by the c$(4\times 2)$ low temperature reconstruction of Si(100)\cite{Wolkow}.\newline
\begin{figure}[htb]
\begin{center}
\includegraphics[width=0.9\textwidth]{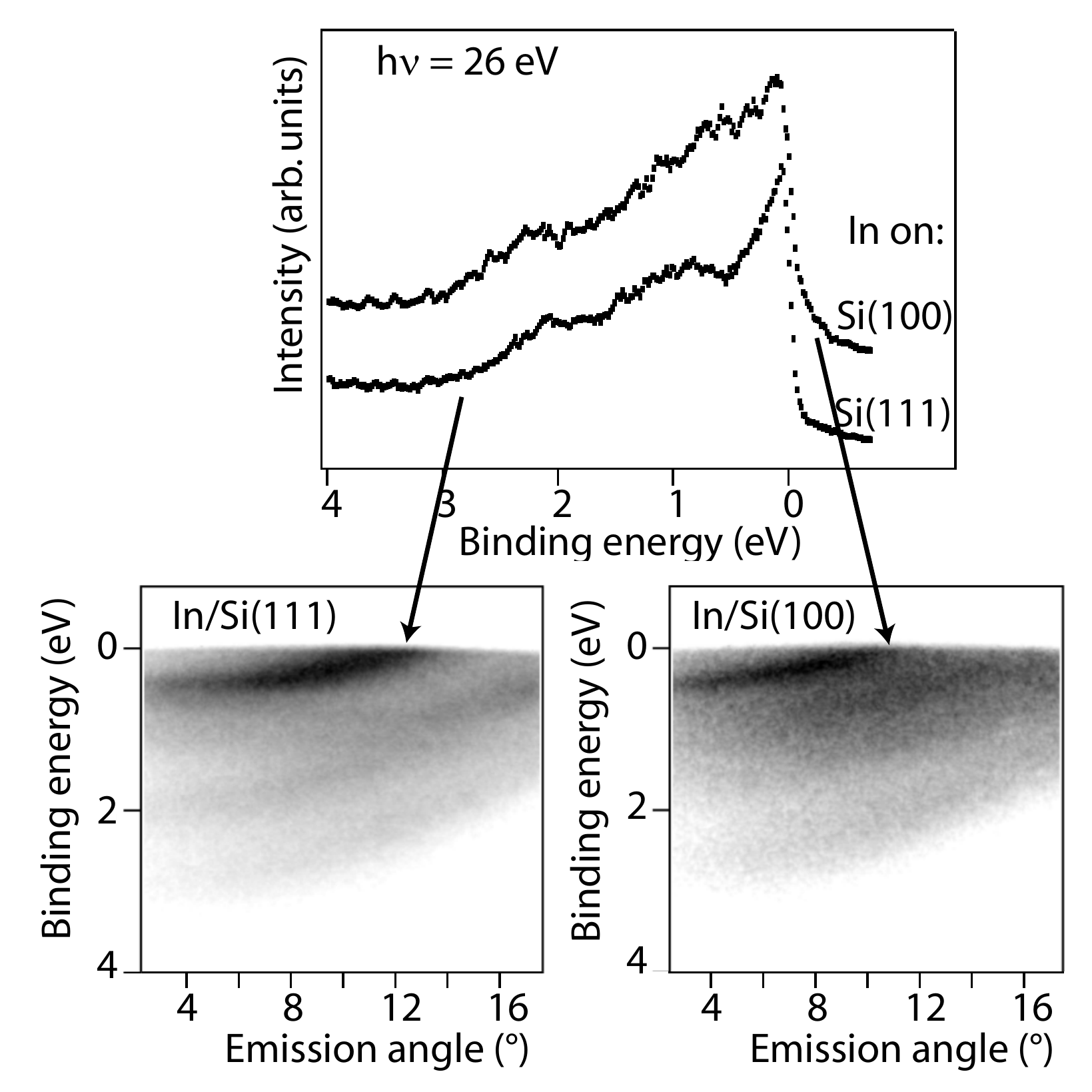} 
\caption{{Off-normal emission images for 10 ML In/Si(111) (left) and In/Si(100) (right) at a photon energy of 26 eV. The arrows indicate where the spectra in the upper part have been extracted. Note that the colour scale of the images has been inverted to obtain a better contrast.}}
\label{Fig5}
\end{center}
\end{figure}
The occurrence of QWS is not influenced by the growth of two different domains of indium overlayers. In the center of the surface Brillouin zone, the $\Gamma-$M direction of domain 1 and the $\Gamma-$K direction of domain 2 exactly overlap. The deviation between the energy bands in the two directions, which can only be probed simultaneously in photoemission, increases for larger k$_{||}$ values. Figure \ref{Fig5} shows a comparison between an energy versus angle image obtained towards the edge of the SBZ in the $\Gamma-$M direction for 10 ML of In on Si(111) (left), and on Si(100) (right). In the image obtained for In/Si(111), the features are significantly sharper than for In/Si(100); this is confirmed by the energy cuts through the image (top), taken at the point where the topmost QWS crosses the Fermi level, as indicated by the arrows. For the In/Si(111) EDC, three individual peaks appear, whereas for In/Si(100) the peaks overlap and can hardly be resolved. Considering the fact that at the given thickness both layers are relaxed towards the \textit{bct} lattice structure, the difference in spectrum quality is likely to be due to the mixture of domains. Because the different domains originate from the two domains of the substrate, suppressing the two-domain growth of indium on Si(100) would seem possible by using the single domain Si(100) substrate with a 4$^\circ$ miscut. 
\begin{figure}[htb]
\begin{center}
\includegraphics[width=0.9\textwidth]{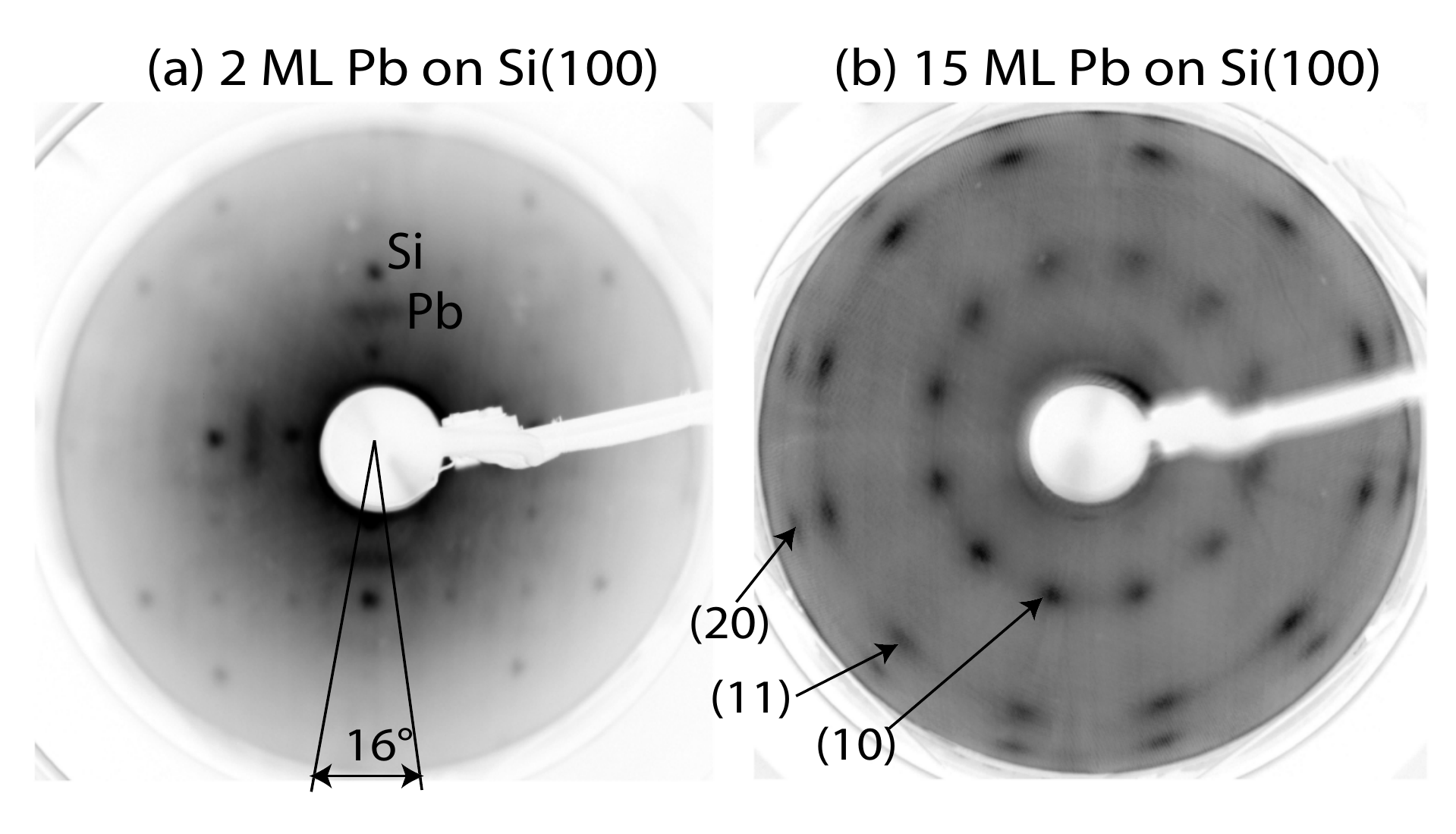} 
\caption{{LEED patterns for (a) 2 ML Pb on Si(100) at 80 eV and (b) 15 ML Pb on Si(100) at 105 eV.}}
\label{Fig6}
\end{center}
\end{figure}
\subsection{Structure of, and QWS in Pb films on Si(100)}
\begin{figure}[htb]
\begin{center}
\includegraphics[width=0.9\textwidth]{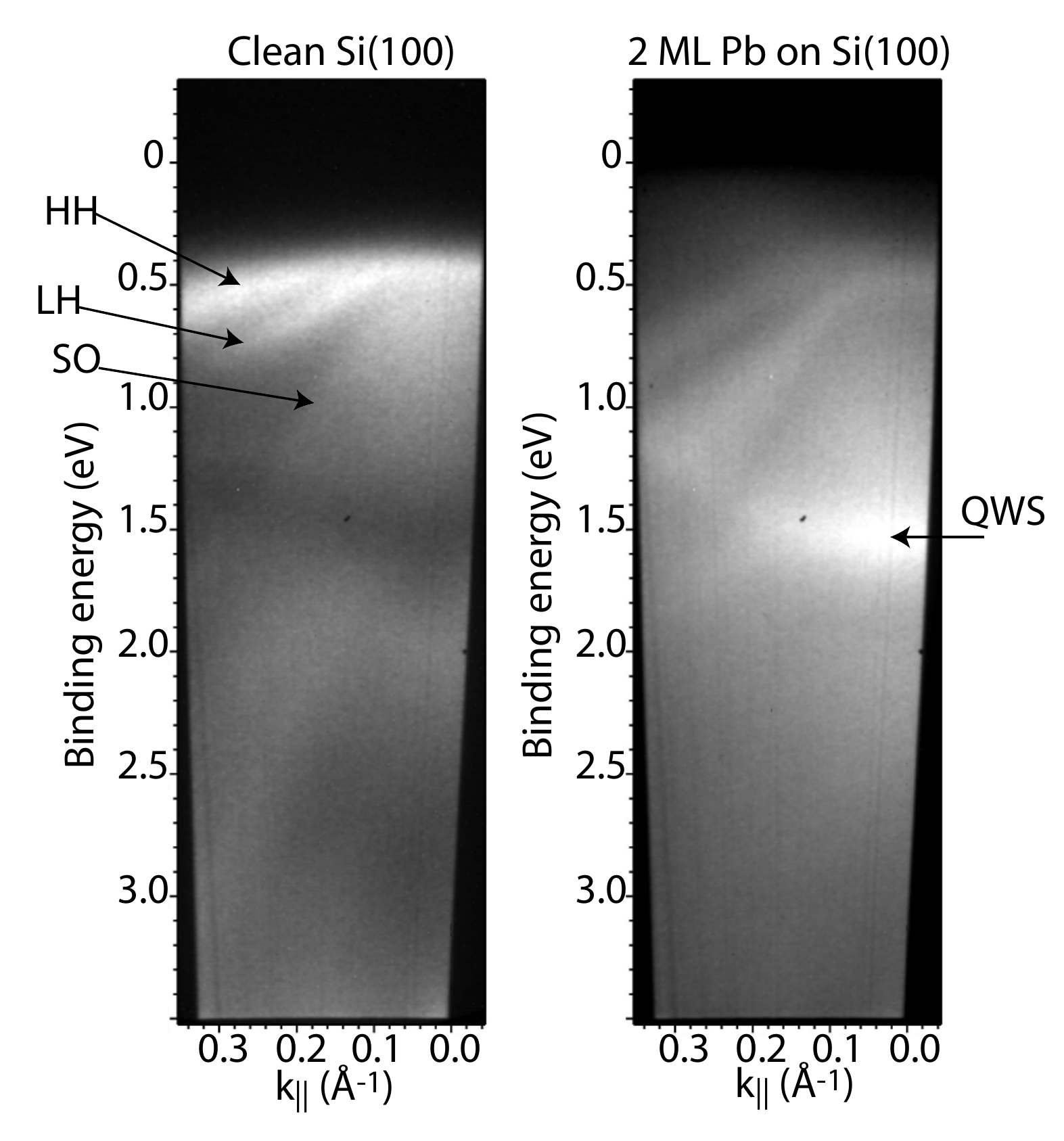} 
\caption{{(left) Energy vs. momentum photoemission intensity image for clean Si(100) at a photon energy of 24 eV with the heavy-hole (HH), light-hole (LH) and split-off (SO) bands indicated. (right) Image obtained with the same parameters for 2 ML of Pb on Si(100), with the QWS indicated.}}
\label{Fig7}
\end{center}
\end{figure}

In this section we focus on a different system (Pb on Si(100)) where the strain induced by the substrate also has a profound influence on the formation of QWS. Figure \ref{Fig6}(a) shows the LEED pattern obtained after the deposition of 2 ML of Pb on Si(100), at an electron energy of 80 eV. On the inside of the sharp spots from the square silicon substrate lattice, an elongated spot can be observed that is not present for clean Si(100). Upon closer inspection we find that this feature is actually composed of two separate spots, which can be assigned to a square lattice rotated by 8$^\circ$ with respect to the Si(100) lattice, suggesting the formation of an epitaxial Pb(100) layer induced by the symmetry of the substrate. This layer should also leave its signature in ARPES, in that a QWS should show up at a binding energy of approximately 1.8 eV, in analogy with our derivation for indium above, and as predicted by DFT calculations\cite{Yu}. Figure \ref{Fig7} shows a comparison between a 5$^\circ$ off-normal emission photoemission image for clean Si(100) and the 2 ML thick Pb(100) film formed on this substrate. The images were obtained with the dispersing angle along the horizontal direction of Figure \ref{Fig6}(a). For the clean substrate, the heavy-hole (HH), light-hole (LH), and split-off (SO) bands\cite{Hermann} are identified and marked accordingly. After Pb deposition, an intense upward dispersing feature is formed, which does not shift with photon energy. Both this characteristic behaviour and the absence of this feature on the clean substrate suggests an assignment to either a quantum well state or an interface state. The good match to DFT results\cite{Yu} indicates at the former. The binding energy of this state is slightly lower than predicted for a free-standing Pb(100) film of 2 ML thickness, which may occur when the substrate influence is not negligible\cite{Dil2}. Emission from the silicon bands remains visible parallel to that from the Pb overlayer. The intensity distribution, however, is different from clean Si, with higher intensities in regions close to the Pb QWS.\newline 
For higher coverages of Pb on Si(100), QWS are no longer observed; only a broad feature, dispersing with photon energy exactly like the Pb bulk band\cite{Horn}, is visible. This feature is comparable to observations for Pb on Cu(111)\cite{Dil2} and Si(111)\cite{Mans} where the films have not been brought into thermal equilibrium, and is due to disorder at the metal/substrate interface. As explained above, the QWS wave function is composed of the rapidly oscillating Bloch wavefunction derived from the Pb atom spacing, modulated by the slowly varying QWS envelope function. This QWS envelope function only develops if a standing electron wave is formed, which in turn requires that the backscattering at the interface is coherent, and that the phase information is preserved while travelling through the layer. The LEED pattern of a 15 ML film in Figure \ref{Fig6}(b) provides a hint why these conditions are not met for Pb layers thicker than 2 ML on Si(100): From the third layer onwards, the Pb no longer grows in a square lattice, but in the hexagonal structure of a Pb(111) layer. Again the two domains of the Si(100) substrate are responsible for the growth of two domains in the Pb film, resulting in 12 spots being observed. This LEED pattern was recorded at an electron energy of 105 eV; at this energy the higher order diffraction spots are also visible (indicated in the image are the (10), (11), and (20) spots for one domain). Due to the 90$^\circ$ rotation between the domains, the (11) and (20) spots from different domains almost overlap. Compared to Pb/Si(111)$(7\times 7)$\cite{Mans}, where QWS are formed, the spots are relatively sharp. Therefore the limiting factor is not the overall crystal structure of the overlayer. For coverages of approximately 4 ML, a combination of the square lattice from Figure \ref{Fig6}(a) and the hexagonal lattice from Figure \ref{Fig6}(b) is observed. This suggests that the 2 ML thick Pb(100) layer survives underneath the Pb(111) film for higher coverages. This change in the orientation of the growth surface apparently inhibits coherent backscattering at the interface, and the formation of QWS, such that only the bulk-like band dispersion appears in the spectra.\newline
We explain the growth mode of Pb on Si(100) by a combination of two effects, namely the orientation of the substrate, and electronic growth effects\cite{Zhang}. The initial growth being in the (100) direction is obviously due to the symmetry of the substrate; however, the formation of a 2 ML thick layer and its survival under thicker films is rationalized by thermodynamic stability considerations. DFT calculations show that the surface energy per unit cell of a free-standing 2 ML thick Pb(100) layer is about one third lower than the surface energy per unit cell of thicker Pb(100) films\cite{Yu}, and almost half of that for a 3 ML thick layer\cite{Wei1}. The formation of a Pb(111) film on a Si(100) substrate involves a considerable amount of strain in the overlayer, which increases the surface energy of the layer. Although the surface energy of Pb(111) layers\cite{Wei2} is much lower than that of Pb(100) layers over the full range studied, the extra energy due to strain shifts the energy minimum for 2 ML thick films in the direction of the Pb(100) structure. Due to the large energy increase for the formation of a 3 ML thick Pb(100) layer, the growth direction then suddenly changes to (111). The 2 ML film is stabilised due to quantum size effects and survives underneath the (111)-like layers. 

\section{Conclusions}
We have characterized the influence of the substrate lattice structure on the formation of quantum well states, through a combination of LEED and photoemission studies and DFT calculations, comparing the systems In on Si(111) and Si(100), and Pb on Si(100). We find that in thin In films, well characterised quantum well states only form after the layer has relaxed into the \textit{bct} lattice structure. Since our DFT calculations show that the In lattice structure has only a small influence on the electronic structure, the occurrence of the QWS may not simply  be related to a transformation of the film structure, but may also depend on the interface structure. For Pb on Si(100) on the other hand, quantum size effects stabilize the 2 ML thick Pb(100) film, but the tendency for energy minimization changes the growth surface for the thicker layers to the (111) direction. No QWS can form in films thicker than 2 ML due to this change in growth direction.\newline 
For both systems a change in lattice structure influences the formation of quantum well states, where the change in growth direction for Pb has a more profound influence as the \textit{fcc} to \textit{bct} relaxation in thin indium films.
\ack
This work was supported as part of the European Science Foundation EUROCORES programme SONS under programme MOL-VIC through funds from the Deutsche Forschungsgemeinschaft, and the EC Sixth Framework programme. J.H.D. acknowledges support by the International Max-Planck Research School "Complex Surfaces in Materials Sciences." We gratefully acknowledge support by G. Reichardt and the BESSY staff.

\end{document}